\documentclass[aps,rmp,twocolumn,showpacs,amsfonts,amssymb,graphicx]{revtex4}
\usepackage{amssymb,amsmath}
\usepackage[dvips]{graphicx}
\usepackage{epsfig}
\usepackage{graphicx}

\begin{document}

\title{A polarizable interatomic force field for TiO$_2$ parameterized using density functional theory}

\author{X. J. Han$^{1,2}$, L. Bergqvist$^{1,3}$, P. H. Dederichs$^1$, H. M\"uller-Krumbhaar$^1$, 
J. K. Christie$^{4,5}$, S. Scandolo$^{5,6}$, and P. Tangney$^{7}$\footnote{Email: p.tangney@imperial.ac.uk}} 

\affiliation{
$^1$ IFF-Theory III, Forschungszentrum J\"ulich, D-52425 J\"ulich, Germany. 
\\
$^2$School of Materials Science and Engineering, Shanghai Jiao Tong University, 200240 Shanghai, P. R. China.
\\
$^3$Department of Physics and Material Science, Uppsala University, Box 530, SE-751 21 Uppsala, Sweden.
\\
$^4$Department of Chemistry, University College London, 20 Gordon Street, London WC1H 0AJ, United Kingdom.
\\
$^5$The Abdus Salam International Centre for Theoretical Physics (ICTP),
     Strada Costiera 11, I-34014 Trieste, Italy. 
\\
$^6$INFM/Democritos National Simulation Center,
     Via Beirut 2-4, I-34014 Trieste, Italy. 
\\
$^7$Department of Physics and Department of Materials, Imperial College London, London SW7 2AZ, United Kingdom. 
}
\date{\today}

\begin{abstract}
We report a classical interatomic force field for TiO$_2$, 
which has been parameterized using density functional theory forces, 
energies, and stresses in the rutile crystal structure. The reliability 
of this new classical potential is tested by evaluating the structural 
properties, equation of state, phonon properties, thermal expansion, 
and some thermodynamic quantities such as entropy, free energy, and 
specific heat under constant volume. The good agreement of our results 
with {\em ab initio} calculations and with experimental data, indicates that 
our force-field describes the atomic interactions of TiO$_2$ in the 
rutile structure very well. The force field can also describe the 
structures of the brookite and anatase crystals with good accuracy.
\end{abstract}
\vspace{0.5cm}

\maketitle
\section{Introduction}
Recently, a bipolar switching phenomenon in TiO$_2$ has been observed \cite{jeong1,jeong2,shima}
and much work has  been done to design superior TiO$_2$-based resistive random 
access memory (RRAM) devices. This, together with many other important 
applications of TiO$_2$, such as white pigment for paints, sunscreen, 
high-efficiency solar cells \cite{regan}, and photosplitting of water to hydrogen \cite{khan,du}, 
has been stimulating a great deal of research interest in atomistic 
simulations of TiO$_2$. Accurate atomistic simulations would provide detailed information 
at the atomic level that might answer some important open questions, such as the 
mechanism of bipolar switching.  

Force fields are the key to accurate atomistic simulations. In the last 
twenty years, many force fields have been built for 
TiO$_2$ \cite{matsui,swamy,ogata,collins,hallil,kerisit,thomas,tetot}. Among these, the partial-charge 
rigid-ion model of Matsui and Akaogi (MA model) \cite{matsui}, 
and the variable-charge Morse stretch (MS-Q) interatomic potential of Swamy\cite{swamy}, 
have been the most successful and popular. Both the MA and MS-Q models can describe 
a series of crystal structures quite well. A common feature of these force fields 
is that they were adapted to reproduce experimental bulk properties, such as 
lattice constants, cohesive energies, bulk moduli and elastic constants. 
An alternative parameterization method consists of fitting force fields to forces, 
energy differences, and stresses extracted from {\em ab initio} calculations
\cite{ercolessi,ts,ts-mgo,brommer}. 
To our knowledge, no such {\em ab initio} force field for TiO$_2$ yet exists.

Having a potential that can accurately model the vibrational properties of TiO$_2$ as 
well as its structures is
very important as they determine its finite-temperature macroscopic properties.
The phonon dispersion curves and density of states for rutile TiO$_2$ 
have been measured by the coherent inelastic scattering of thermal neutrons along 
principal symmetry directions of the Brillouin zone \cite{traylor}. The $\Gamma$-point phonons have 
been measured by Raman \cite{porto} and infrared \cite{spitzer} spectroscopy. In addition to these 
experimental determinations, phonon frequencies have been calculated {\em ab initio}\cite{montanari,lee,sikora}. 
However, little attention has been paid to the determination of phonon properties from 
a force field for TiO$_2$, even though it is a very demanding test of a force field's accuracy. 
As has been demonstrated for MgO\cite{ts-mgo,aguado}, {\em ab initio}-based classical force fields 
have the potential to give a 
very good description of phonon properties. 
TiO$_2$ has several competing lattice structures and is a more complex oxide than MgO. It is rather interesting, 
therefore, to see how well such a force field can predict its vibrational properties.

In this work, we present a polarizable classical interatomic force field for TiO$_2$, parameterized
from {\em ab initio} calculations. We test its 
accuracy by comparing its predictions of structural properties, 
equations of state, phonon frequencies, and the temperature dependences of lattice parameters, entropy, 
free energy, and specific heat under constant volume in the rutile phase with experimental data
and with {\em ab initio} calculations. We also calculate the structures and relative energies
of the brookite and anatase phases to get an indication of the force field's transferability.

\section{Force field}
The potential energy function $U = U^{\textrm{sr}}+U^{\textrm{es}}$
that we use to describe interactions between ions
consists of a pairwise term $U^{\textrm{sr}}$ describing short-range non-electrostatic
interactions and an electrostatic term $U^{\textrm{es}}$ 
describing dipole induction and the electrostatic 
interactions of the charges and induced dipole moments of the ions.

For the short-range interaction energy we use the pairwise Morse-stretch form
\begin{eqnarray}
U^{\textrm{sr}} 
& = & \sum_{i>j}D_{ij} \big[ e^{\gamma_{ij} [1-(r_{ij}/r_{ij}^0)]}
-2e^{(\gamma_{ij}/2) [1-(r_{ij}/r_{ij}^0)]}\big]
\end{eqnarray}
where $r_{ij}=|{\bf r}_i-{\bf r}_j|$ is the distance 
between nearby ions $i$ and $j$ and 
$D_{ij}$, $\gamma_{ij}$, and $r_{ij}^0$ are 
parameters that are specific to the pair of species of ions
$i$ and $j$. This potential is truncated at a radius of $18.0$ a.u.

The total electrostatic contribution to the energy of the system includes charge-charge, 
charge-dipole, and dipole-dipole interaction terms: 
\begin{eqnarray}
U^{\textrm{es}} & = & \frac{1}{4\pi\epsilon_0} \sum_{i>j}\bigg[ \frac{q_i q_j}{r_{ij}} + \sum_{\gamma} \nabla_\gamma 
\bigg(\frac{1}{r_{ij}}\bigg) \big( p_i^\gamma q_j-q_i p_{j}^\gamma\big)  \nonumber \\
& - & \sum_{\gamma,\beta}
\nabla_\beta \nabla_\gamma \bigg(\frac{1}{r_{ij}}\bigg) p_i^\gamma p_j^\beta \bigg] 
+ \sum_{i,\gamma} \frac{(p_i^\gamma)^2}{2\alpha_i}
\end{eqnarray}
where $q_i$ is the charge of ion $i$, $p_i^\gamma$ is the $\gamma^\textrm{th}$ cartesian component of the
dipole moment of ion $i$, $\alpha_i$ is its polarisability, and $\nabla_\gamma = \partial/\partial r^\gamma_{ij}$.
The final term on the right hand side of this equation is a sum of the self energies of the ions.
The self energy of an ion is the internal energy cost of inducing a dipole on it.

There are two mechanisms for the induction of ionic dipole moments. The first is via the short-range
repulsive forces between ions which can distort an ion's electron cloud thereby giving it a dipole moment.
The second is the induction of a dipole moment on an ion by the electric field arising from the charges and 
the dipole moments of all other ions.

To model the induction by short-range repulsive forces 
we follow the approach of Madden {\em et al.}\cite{wilson-madden1,rowley}.
In their approach, the contribution of the short-range forces to dipole moments is given by
\begin{equation}
{\bf p_i^{\text{sr}}} = \alpha_i \sum_{j\neq i} \frac{q_j {\bf r}_{ij}}{r_{ij}^3}f_{ij}(r_{ij})
\end{equation}
where
\begin{equation}
f_{ij} (r_{ij}) = c_{ij} \sum_{k=0}^4 \frac{(b_{ij}r_{ij})^k}{k!} e^{-b_{ij}r_{ij}}
\end{equation}
and $b_{ij}$ and $c_{ij}$ are model parameters.

The electrostatically-induced dipole moments are more difficult to calculate because they 
are all interdependent due to their contributions to, and dependences on, the electric field.
At every time step we find the dipole moments of the ions by iterating to self-consistency the equation
\begin{equation}
{\bf p}_i^m = \alpha_{i} {\bf E}({\bf r}_i; \{{\bf p}_j^{m-1}\}_{j\neq i}; \{ {\bf r}_j\}_{j\neq i}) 
+ {\bf p}_i^{\text{sr}}
\end{equation}
where ${\bf p}_i^m$ is the dipole moment on ion $i$ at iteration $m$ of the 
self-consistent cycle; ${\bf E}({\bf r}_i)$ 
is the electric field at position ${\bf r}_i$. 

We use Ewald summation to calculate electrostatic energies, forces, stresses, and electric fields.

Our model depends on a set of parameters $\{\eta_n\}$ comprising the Morse-Stretch parameters $D$, $\gamma$, 
and $r^0$, the parameters $b$ and $c$ of the short-range dipole model, 
and the charges $q_\textrm{\tiny Ti},q_\textrm{\tiny O}$ and 
polarisabilities $\alpha_\textrm{\tiny Ti},\alpha_\textrm{\tiny O}$ of 
the ions. In section \ref{section:parameters} we describe how these parameters are determined.

\section{Potential parameterization} \label{section:parameters}
\subsection{{\em Ab initio} molecular dynamics}
Total energies, stresses, and forces are computed from {\em ab initio} 
simulations based on density functional theory (DFT). {\em Ab initio} molecular dynamics (MD) simulations 
employing the projector augmented wave method (PAW) were performed in the NVT ensemble  
under periodic boundary conditions using the VASP 
simulation package \cite{vasp1,vasp2,vasp3}. Only the Ti (3d,4s) and O (2s,2p) electron states were treated as 
valence states, however, tests in which the Ti (3s, 3p) semicore states were included yielded similar results
for the energy differences between crystalline phases and the phonon frequencies of rutile TiO$_2$.
The local-density approximation (LDA) to the 
exchange-correlation energy was used. The velocity Verlet algorithm \cite{verlet}
with a time step of $1$ fs was adopted to solve the equations of motion, 
and temperature was controlled using a Nose-Hoover thermostat\cite{nose}. 
We used a $2\times 2 \times 3$ supercell that contained $72$ atoms and only 
the $\Gamma$-point was used to sample the Brillouin zone. 
A kinetic energy cutoff of $180$ eV was used in the plane-wave expansion of the 
wave functions during the MD simulations. The goal of these {\em ab initio} MD 
simulations was simply to produce some reasonable atomic configurations to serve as
representative snapshots at finite temperature. These configurations were then 
used to perform higher-precision DFT calculations of forces, stresses, and 
energies for use in the force-field parameterization.
\subsection{Fitting procedure for the potential}
We fit our potential to DFT calculations in the rutile crystal structure, however, 
to broaden the range of different environments for the ions, two temperatures 
are considered in the potential fitting: 300K and 1000K. Furthermore, at each 
temperature, configurations at more than one volume were chosen. For 300K eight configurations
were used. Six of these configurations are taken  
from an {\em ab initio} MD simulation at the zero temperature equilibrium volume. 
The other two were obtained from simulations in which the volume was expanded by 3\%. 
At 1000K, we used 3 sets of data in total, each containing four configurations. 
These three sets are obtained from MD simulations at the zero-temperature 
equilibrium volume, and at volumes expanded by 3\% and by 9\%.
To minimize correlations between configurations, successive snapshots were separated from 
one another by 2 ps of {\em ab initio} MD. For each snapshot, we used VASP to calculate the 
forces, energies and stresses.  To obtain these quantities with high precision, the Brillouin zone 
was sampled using a $2\times 2 \times 2$ Monkhorst-Pack $k-$point mesh and, to 
converge the forces and the stress tensors properly, a very high wavefunction 
cutoff energy of $1500$ eV was used. 

The potential parameters were obtained by minimizing the cost function
\begin{equation}
\Gamma(\{\eta_n\}) = w_{f}\Delta F 
+ w_{s}\Delta S
+w_{e}\Delta E
\end{equation}
with respect to the parameters $\{\eta_n\}$ where 
\begin{eqnarray}
\Delta F & = & \frac{\sqrt{\sum_{k=1}^{N_{c}}\sum_{I=1}^{N}\sum_{\alpha} |F_{cl,I}^{\alpha}(\{\eta_n\})
-F_{ai,I}^{\alpha}|^{2}}}
{\sqrt{\sum_{k=1}^{n_{c}}
\sum_{I=1}^{N}\sum_{\alpha} (F_{ai,I}^{\alpha})^{2}}}   \\
\Delta S & = & 
\frac{\sqrt{\sum_{k=1}^{N_{c}}\sum_{\alpha,\beta} |S_{cl}^{\alpha\beta}(\{\eta_n\})-S_{ai}^{\alpha\beta}|^{2}}}
{3B\sqrt{n_{c}}}  \\ 
\Delta E & = &
\frac{ \sqrt{\sum_{k,l}^{N_{c}} ((U_{k}^{cl}-U_{l}^{cl})-(U_{k}^{ai}-U_{l}^{ai}))^{2}}}
{\sqrt{\sum_{k,l}^{n_{c}}(U_{k}^{ai}-U_{l}^{ai})^{2}}} 
\end{eqnarray}
$F_{cl,i}^\alpha$ is the $\alpha$-component of the force on ion $i$ calculated 
with the classical potential; $F_{ai,i}^\alpha$ is 
the $\alpha$-component of the force on ion $i$ calculated {\em ab initio}; 
$S_{cl}^{\alpha\beta}$  
is the stress tensor component 
calculated with the  classical potential; $S_{ai}^{\alpha\beta}$ is the 
stress tensor component calculated {\em ab initio}; 
$N_c$ denotes the number of snapshot configurations used in the fit; $B$ is the bulk modulus; 
$U_{cl}^k$ is the potential energy of configuration $k$ for the classical potential, 
and $U_{ai}^k$ is its energy  
calculated {\em ab initio}. $w_f$, $w_s$, and $w_e$ are weights that quantify the relative importances to the
fitting cost function of forces, stresses, and energies, respectively. 
In this work, $w_f =1.0$, $w_s =0.5$, and $w_e =0.01$. 
These numbers are somewhat arbitrary as long as the energy differences are given a relatively small weight
because we only have one energy per configuration.  We tried different weights and found no significant 
difference in the final values of $\Delta F$, $\Delta S$, and $\Delta E$.
  Minimization of $\Gamma(\{\eta_n\})$ is performed by a combination of simulated annealing \cite{simann} and 
Powell minimization\cite{numrec}. The former is used to provide an initial parameter set which brings the 
cost function to a basin in the surface defined by $\Gamma(\{\eta_n\})$ in $\eta-$space. 
Minimization is then completed using Powell's method. 

We assume that DFT provides accurate atomic forces and energies.
Therefore if, at the global minimum of $\Gamma$ in $\eta-$space, $\Delta F$ remains large
it means that the potential model does not contain the ingredients necessary to 
fit the DFT potential energy surface closely. The model is unphysical or incomplete
and the resulting force-field is 
unlikely to produce results that are in good agreement with experiment. 
Agreement with experiment can still be found by fitting to experimental data, but when agreement 
is not attributable to accurate interatomic forces, it is difficult to place confidence in the force-field
for calculations that can not be directly checked by comparison with experiment. However, when reliable experimental data 
is available, simulations are of limited value anyway. For this reason we deem it important for a force-field to be
able to fit DFT forces closely and the closeness of the fit achieved in the parameterisation is itself
a test of the quality of the potential.

\section{Results}
Following the fitting procedure described in the previous section, we obtained 
the force-field parameters for rutile TiO$_2$, as listed in table \ref{table:one}. 
In this fitting, $\Delta F = 0.202$, $\Delta S=0.009$, and $\Delta E=0.046$. 
Put another way, the root-mean-squared error in the forces is $20.2\%$ of the
root-mean-squared force.

As a comparison we have checked how close the MA model's forces are to the {\em ab initio} ones.
We find that the root-mean-squared difference between the MA forces and the DFT forces is $\sim 88\%$
of the root-mean-squared DFT forces 
($\Delta F \approx 0.88$, $\Delta S \approx 0.02$, $\Delta E \approx 0.63$).
The MA model uses the pairwise Born-Mayer potential form and
by minimising $\Gamma$ with respect to the parameters of the Born-Mayer 
potential we have been able to achieve a best fit of 
$\Delta F \approx 0.48$, $\Delta S \approx 0.01$, $\Delta E \approx 0.10$. However, this
resulted in a disimprovement in the description of the three crystal structures with respect
to the MA potential. As one might expect, and as was found previously for silica\cite{ts}, 
when the fit to {\em ab initio}
calculations is poor, the ability of a potential to reproduce experimental results does
not correlate with the quality of the fit.
Our tests strongly suggest that the Born-Mayer form provides a poor microscopic description of atomic forces in 
TiO$_2$.

\subsection{Crystal structures}
In order to check the reliability and transferability of the new force field, 
we computed the structural properties of three different crystalline polymorphs 
of TiO$_2$, namely the rutile, anatase, and brookite structures. Structural 
optimizations for these three crystals were performed via the steepest-descent 
method at zero pressure. The resulting lattice parameters are presented in Table \ref{table:two} 
and compared to the {\em ab initio} data calculated with the VASP package and to experimental data\cite{abrahams,burdett,meagher}, 
both at room temperature and (for rutile and anatase) at $15$ K.  
Both {\em ab initio} calculations and the new force field reproduce experimental data quite well 
and the accuracy of the force field is comparable to that of DFT. We stress that the 
force field in this work is only fit to {\em ab initio} data from calculations in the rutile 
crystal structure. Therefore, its ability to reproduce the structures of anatase and brookite 
is an indication that it may be reasonably transferable between different structures. 
However, apart from the handful of structural and energetic properties presented in this 
paper, we have not tested the force field in these phases.

\begin{table}
\begin{ruledtabular}
\begin{tabular}{c|c|c|c}
Parameters  & O-O & Ti-O & Ti-Ti \\ \hline 
$D$      & $1.5865\times10^{-3}$& $1.7668\times10^{-3}$ & $5.0276 \times 10^{-3}$ \\
$\gamma$ & 9.6370        & 12.2332   &5.9069  \\
$r^0$      & 6.5150        & 4.7678    & 8.1380\\
$b$      & 1.5368        & 4.8187    & 1.8246\\
$c$      & 2.9216        & -122.1489 & -0.7918\\
$\alpha$    & 2.9314        &           & 10.2739\\
$q$         & -1.1045       &           & 2.209\\ 
\end{tabular}
\end{ruledtabular}
\caption{The fitted interatomic force-field parameters (in atomic units).}
\label{table:one}
\end{table}

\begin{table*}
\begin{ruledtabular}
\begin{tabular}{c|ccc|ccc|ccc}
 & \multicolumn{3}{c|}{a (\AA)} & \multicolumn{3}{c|}{b (\AA)} & \multicolumn{3}{c}{c (\AA)} \\ 
  & Exp. & DFT-LDA & POT & Exp. & DFT-LDA & POT & Exp. & DFT-LDA & POT \\ \hline
Rutile & & & & & & & & & \\
$\sim$ 0 K  & 4.587\cite{burdett} &4.571 & 4.494 & & & & 2.959\cite{burdett} & 2.925 & 3.015  \\
$\sim$ 300 K& 4.593\cite{burdett} &      & 4.504 & & & & 2.954\cite{burdett} &       & 3.023  \\ \hline
Anatase& & & & & & & & & \\
$\sim$ 0 K   & 3.782\cite{burdett} & 3.821 & 3.790 & & & & 9.512\cite{burdett} & 9.694 & 9.292 \\
$\sim$ 300 K& 3.785\cite{burdett} &      &       & & & & 9.502\cite{burdett} &       &       \\ \hline
Brookite&&&&&&&&& \\
$\sim$ 0 K &                    &9.1197& 9.095 &  & 5.415 & 5.399 &  & 5.103 & 5.131 \\
$\sim$ 300 K & 9.174\cite{meagher} &  & & 5.449\cite{meagher} & & & 5.138\cite{meagher} & & \\ 

\end{tabular}
\end{ruledtabular}
\caption{Lattice parameters for three optimised structures of TiO$_2$, together with experimental 
data and results from {\em ab initio} calculations. DFT-LDA represents {\em ab initio} calculations with 
the VASP package within the local density approximation, and POT the new force field, obtained by 
fitting to {\em ab initio} data in the rutile structure. 
Data at two temperatures are presented, where available: at close
 to 0 K (15 K in the experiment of Ref.~\onlinecite{burdett}, 0 K for simulation) and at room temperature.} 
\label{table:two}
\end{table*}

\begin{table*}
\begin{ruledtabular}
\begin{tabular}{c|cc|ccc|cc}
  & \multicolumn{2}{c|}{V$_0$/TiO$_2$ (\AA$^3$)} & \multicolumn{3}{c|}{B$_0$ (GPa)} 
& \multicolumn{2}{c}{B$^*_0$ } \\ 
  & DFT-LDA & POT & Exp. & DFT-LDA & POT & DFT-LDA & POT \\ \hline 
Rutile & 30.62 & 30.46 & 211$\pm$ 7\cite{ming} & 252 & 241 & 5.37 & 5.07  \\
   &        &         &  230$\pm$ 20\cite{gerward} & & & & \\      
Anatase & 33.58 & 33.37 & 178\cite{swamy} & 188 & 171 & 1.75 & 2.34 \\
Brookite & 31.55 & 31.50 & no data & 230 & 196 & 3.87 & 3.14  \\ 
\end{tabular}
\end{ruledtabular}
\caption{Birch-Murnaghan EOS parameters for different structures of TiO$_2$, together with the experimental data available. 
Experimental data refers to room temperature, while theoretical values are obtained at $0$ K.}
\label{table:four}
\end{table*}

\begin{table*}
\begin{ruledtabular}
\begin{tabular}{c|cccccc}
 & Experiment & DFT-LDA & DFT-GGA & POT & MA \\ \hline
E$_{\text{anatase}}$-E$_{\text{rutile}}$ & 35.0\cite{mitsuhashi}, 53.8\cite{navrotsky} & -12.0 & -81.2 & 425.9 & 301.6\\
E$_{\text{brookite}}$-E$_{\text{rutile}}$ & 7.7\cite{mitsuhashi} & -17.5 & -39.9 & 212.6 & 179.3 \\ 
\end{tabular}
\end{ruledtabular}
\caption{Energy differences in meV/TiO$_2$ between the anatase and brookite crystal structures and the 
rutile crystal structure. The experiments of reference \onlinecite{navrotsky} and the calculations 
with the MA model\cite{matsui} were performed at room temperature, while the more recent
experiments of reference \onlinecite{mitsuhashi} 
were performed at $971$ K. The experiments measured enthalpies of transformation whereas
all the  DFT and force-field calculations that we report are energy differences at $0$ K.
 }
\label{table:three}
\end{table*}

\subsection{Equations of state}
\begin{figure}
\epsfig{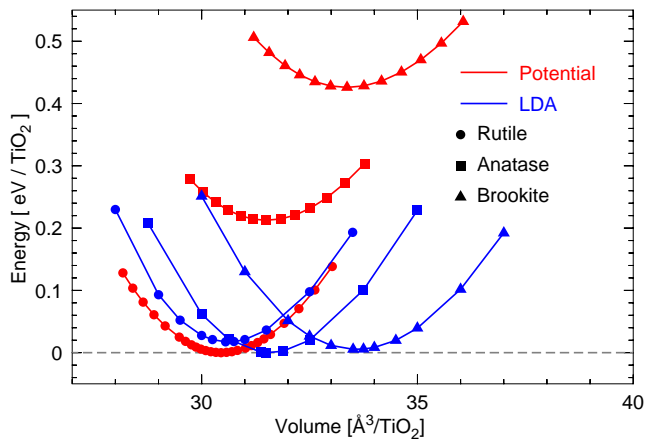}
\caption{Volume dependence of the energy of TiO$_2$ in the rutile, brookite, and anatase crystal structures for our potential (red lines) and from 
LDA {\em ab initio} calculations. Energies are expressed relative to the equilibrium energy of the most stable phase.}
\label{fig:volene}
\end{figure}

The energy as a function of volume for all three phases has been 
calculated with our potential and the results are illustrated in Fig. \ref{fig:volene}.
Structural relaxations were performed at each volume. 
The results in red are from the new force field, while those in blue are 
from our density functional theory calculations within the local density approximation (DFT-LDA). 
In each case, we set the energy of the most 
stable phase in equilibrium to zero. We can see that the DFT-LDA calculations 
and the new force field give similar equilibrium volumes for the three crystals (Table \ref{table:four}). 
However, it is clear from Fig. \ref{fig:volene} and Table \ref{table:three},
that the energy sequences obtained from DFT-LDA calculations and the 
new force field are different and, furthermore, that the the magnitudes of the
energy differences between the three structures differ substantially between 
DFT-LDA and our force field. In DFT-LDA calculations, the energy differences 
between the three structures are small, 
and $E_{\text{brookite}} < E_{\text{anatase}} < E_{\text{rutile}}$. However, for 
the new potential, the rutile structure is the ground state, 
and $E_{\text{rutile}} < E_{\text{brookite}} < E_{\text{anatase}}$. 
While this energy ordering is consistent with the experimental results 
\cite{mitsuhashi,navrotsky} the magnitudes 
of the energy differences are at least an order of magnitude too large. The DFT-LDA estimates of the 
magnitudes of the energy differences, both here and in previous work\cite{labat}, are in much better agreement 
with experiment, despite the fact that they predict the wrong sequence of structures. 
We have also calculated the equilibrium energy differences within a generalized 
gradient approximation (GGA)\cite{pbe} to the exchange-correlation energy. 
The GGA differences in energy between anatase and rutile and between brookite and rutile are also
presented in Table \ref{table:three}. The magnitudes of these energy differences are larger than those
of LDA, but still a factor of $\sim 5$ smaller than we find with our force field.
We are unaware of 
compelling reasons why LDA should outperform GGA for TiO$_2$. Therefore, it seems reasonable 
to consider differences between LDA and GGA as lower bounds on the error bars associated 
with our approximation to the exchange-correlation energy. The free-energy differences
calculated with the MA model at room temperature are similar in magnitude (but slightly smaller)
to the $0$ K energy differences calculated with our force field and the MA model's energy ordering of 
the three crystal structures is also consistent with experiment.
Although our force field produces energy differences that are much too large, it is
at least gratifying 
that the crystal structures have the correct energy sequence.

The relationship between energy and volume can be described 
by the Birch-Murnaghan equation of state\cite{birch} 
\begin{eqnarray}
E(V) - E_0 & = & \frac{9V_0B_0}{16}\Bigg \{ \Bigg[ 
\bigg(\frac{V_0}{V}\bigg)^{\frac{2}{3}}-1\Bigg]^3 B_0^* \nonumber \\
& + & \Bigg[ \bigg(\frac{V_0}{V}\bigg)^{\frac{2}{3}}-1\Bigg]^2
 \Bigg[ 6-4\bigg(\frac{V_0}{V}\bigg)^{\frac{2}{3}}\Bigg] \Bigg\}
\end{eqnarray}
where $V$ is the volume, $V_0$ the equilibrium volume, $E_0$ the equilibrium energy, $B_0$ the bulk modulus, and $B_0^*$ 
the pressure derivative of the bulk modulus. The parameters for the equation of state are listed in 
Table \ref{table:four}, together with experimental data \cite{ming,gerward,swamy}. 
The results from the {\em ab initio} calculations, the new classical potential, and the experimental data are all 
in good agreement, however, it is important to note that the simulation results refer to zero temperature while the 
experiments were performed at room temperature.

\subsection{Vibrational properties of rutile}
\begin{figure}
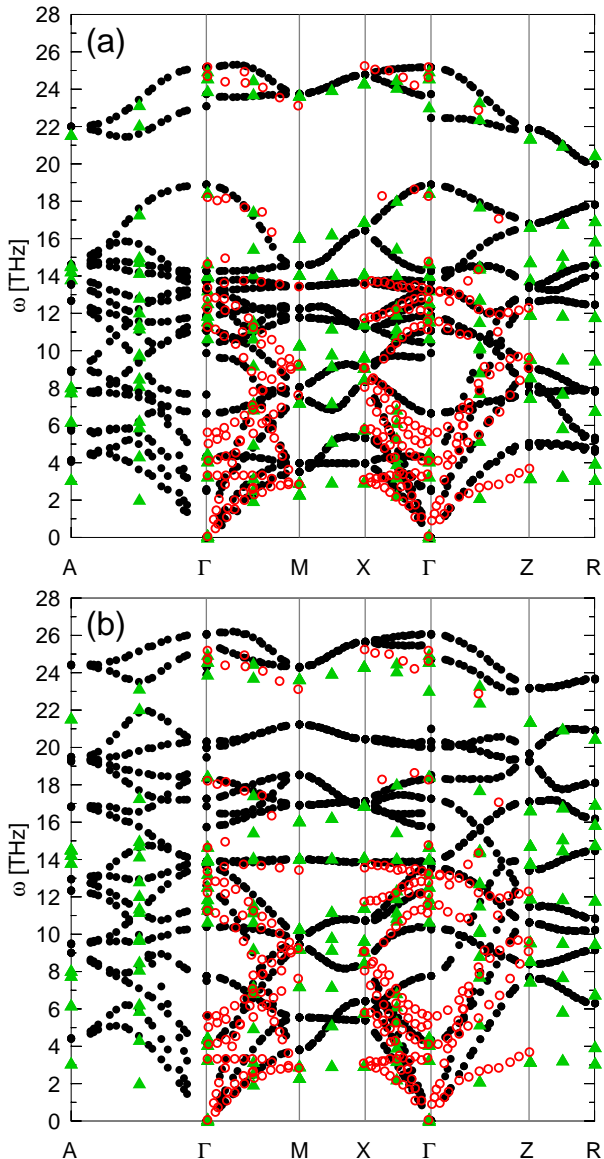

\epsfig{figure=figure2a.ps,width=8.0cm}
\epsfig{figure=figure2b.ps,width=8.0cm}
\caption{Phonon dispersion curves of TiO$_2$ in the rutile crystal structure from different methods.
(a)Results from our new force field (black dots) compared to experiment\cite{traylor} (red circles) and {\em ab initio} (DFT-LDA) calculations (green triangles).
(b)Results from the MA force field (black dots) compared to experiment\cite{traylor} (red circles) and {\em ab initio} (DFT-LDA) calculations (green triangles).}
\label{fig:phonon}
\end{figure}

To have a further test of the new force field, 
the phonons of rutile TiO$_2$ are calculated with the 
small displacement method implemented by Togo's FROPHO package \cite{togo}. 
Figure \ref{fig:phonon}(a) compares phonons calculated with our classical 
potential (black dots) with those calculated from DFT-LDA (green triangles) and, for some branches
and symmetry directions, with inelastic neutron data at room temperature taken
from reference \onlinecite{traylor}.  Figure \ref{fig:phonon}(b) shows the same {\em ab initio} and experimental
data, but the black dots are now from calculations using the MA model\cite{matsui}.
We include in both these plots only those frequencies calculated for phonons whose wavelengths
are commensurate with our simulation cells, i.e., no interpolation has been performed to 
infer frequencies at different wavevectors. Simulations with the classical potentials were performed with
three different simulation supercells whose sizes, 
in multiples of the primitive $6$-atom unit cell, were
$14\times14\times14$ ($16464$ atoms), 
$12\times12\times 12$ ($10368$ atoms), and
$10\times10\times10$ ($6000$ atoms).
Our DFT calculations were performed in a $4\times4\times4$ ($384$ atoms) supercell.

In our calculations we have corrected longitudinal optical frequencies in the long wavelength
limit (i.e. ${\bf k} \rightarrow 0$) to account for the effects of the long-range electric fields
that these modes induce \cite{born-huang,togo}.

The first thing to note from Figures \ref{fig:phonon}(a) and \ref{fig:phonon}(b) is that both classical
potentials struggle to reproduce the dispersion of the acoustic modes close to the zone boundaries.
This can be seen most clearly along the $\Gamma\rightarrow A$ and $\Gamma \rightarrow Z$ symmetry
directions. However, what is also clear is that our force field greatly improves on the MA model
in this respect.
For the high-frequency modes, between $\sim 15$ THz and $\sim 25$ THz, 
our model is in very good agreement with experiment and with DFT. The MA model's description of 
the phonon spectrum at these frequencies is very poor by comparison.
Overall, it is clear that our potential provides a description of the phonon
spectrum that is much better than that provided by the MA model. It is also clear, however, that
there is room for improvement, presumably by improving the functional form of the potential.

\subsection{``Electronic'' properties}
To correct the LO frequencies at the $\Gamma$-point, it has been necessary to calculate 
the high-frequency dielectric tensor  ${\bf \varepsilon}_\infty$, which is diagonal, 
and the Born effective charge tensor ${\bf Z}^*$. 
We have calculated ${\bf \varepsilon}_\infty$ 
and ${\bf Z}^*$ directly by calculating the polarization responses to small electric 
fields\cite{liang} and to small displacements of the atoms, respectively. 
In the rutile crystal structure we find the components of ${\bf \varepsilon}_\infty$ 
perpendicular and parallel to the c-axis to be $\varepsilon_{\perp} =2.58$ and 
$\varepsilon_\parallel = 2.75$, 
respectively, which is in poor agreement with those calculated {\em ab initio}\cite{rignanese}
($\varepsilon_\perp= 7.49$, $\varepsilon_\parallel= 8.57$).  
The Born effective charge tensor of rutile has only 
three independent non-zero components for each atom. We calculate them to 
be 
$Z^*_{xx}(\text{Ti}) = Z^*_{yy}(\text{Ti}) = 3.428$, $Z^*_{xy}(\text{Ti}) = Z^*_{yx}(\text{Ti}) = 0.422$, 
$Z^*_{zz}(\text{Ti}) = 3.789$ 
and $Z^*_{xx}(\text{O}) = Z^*_{yy}(\text{O}) = -1.714$, $Z^*_{xy}(\text{O}) = Z^*_{yx}(\text{O}) = -1.044$, 
$Z^*_{zz}(\text{O}) = -1.895$. 
Again, this is in poor agreement with the {\em ab initio} results\cite{rignanese} which are 
$Z^*_{xx}(\text{Ti}) = Z^*_{yy}(\text{Ti}) = 6.36$, $Z^*_{xy}(\text{Ti}) = Z^*_{yx}(\text{Ti}) = 1.00$, 
$Z^*_{zz}(\text{Ti}) = 7.52$ 
and $Z^*_{xx}(\text{O}) = Z^*_{yy}(\text{O}) = -3.18$, $Z^*_{xy}(\text{O}) = Z^*_{yx}(\text{O}) = -1.81$, 
$Z^*_{zz}(\text{O}) = -3.76$. 
Our potential is not intended to give a good description of the electronic properties of the system,
so this disagreement is no cause for concern.  We intend, instead, an accurate representation of the 
interatomic forces.
Both $Z^*$ and $\varepsilon_\infty$ are derivatives of the 
polarization field, and are therefore only meaningful on length scales 
that are large compared to the primitive unit cell of the crystal. 
At large distances, forces between atoms (labeled ‘1’ and ‘2’) are 
proportional to $Z^*_1 Z^*_2/\varepsilon_\infty$, and so this is the 
relevant quantity if one is interested assessing the quality of these response functions'
contributions to interatomic forces.
Therefore, we look at the components of the screened effective charge tensor 
$Z^*_{\alpha\beta}/\sqrt{\varepsilon_{\gamma\delta}}$. The symmetry
of the crystal is such that, for each atomic species, 
there are only three non-unique non-zero components of the screened effective
charge tensor:
$Z^*_{xx}/\sqrt{\varepsilon_\perp} = Z^*_{yy}/\sqrt{\varepsilon_\perp}$, 
$Z^*_{xy}/\sqrt{\varepsilon_\perp} = Z^*_{yx}/\sqrt{\varepsilon_\perp}$, and
$Z^*_{zz}/\sqrt{\varepsilon_\parallel}$.  
\begin{table*}
\begin{ruledtabular}
\begin{tabular}{c|cc|cc|}
  & \multicolumn{2}{c|}{DFT} & \multicolumn{2}{c|}{POT} \\ 
  & Ti & O & Ti & O \\ \hline 
$Z^*_{xx}/\sqrt{\varepsilon_\perp} = Z^*_{yy}/\sqrt{\varepsilon_\perp}$ & 2.32 & -1.16  & 2.13 & -1.07  \\
$Z^*_{xy}/\sqrt{\varepsilon_\perp} = Z^*_{yx}/\sqrt{\varepsilon_\perp}$ & 0.37 & -0.66  & 0.26 & -0.65  \\
$Z^*_{zz}/\sqrt{\varepsilon_\parallel}$ & 2.57 & -1.28 & 2.28 & -1.14
\end{tabular}
\end{ruledtabular}
\caption{Components of the screened effective charge tensor calculated with density functional
theory (DFT) and with our new potential (POT)}
\label{table:z}
\end{table*}

We have calculated these quantities, which determine long-range forces, with our force field and, 
in Table \ref{table:z},
we compare our results to those
obtained from DFT. 
The agreement between our potential and DFT is good, 
therefore our potential should describe long-range forces reasonably well.

The best-fit ionic polarisabilities ( $\alpha_{\text{O}}=2.9$ a.u., $\alpha_{\text{Ti}}=10.3$ a.u.) 
(Table \ref{table:one})
provided by our parameterization process are unusual and worthy of comment. 
The polarisability of the oxygen ion is much lower than has been found by force fitting 
for other oxides such as MgO\cite{ts-mgo} and SiO$_2$\cite{ts},  or from electronic structure 
simulations\cite{aguado}. Typically, $\alpha_{\text{O}}$ is between $5$ and $15$ a.u.. 
However, most surprising is the large polarizability of the Ti ions. 
One would expect a small polarizability for a cation, particularly one with such a high positive charge.

By making atoms polarisable, we include the response of electrons in a phenomenological way. 
However, we only include one response mechanism. Others, such as higher-order polarisabilities 
and charge-transfer between ions, are certain to exist to some degree. Furthermore, we are 
using a fully-ionic model while covalent effects might be important. It seems likely that 
including one or more of these effects would allow our parameterizer to achieve a closer 
fit to the {\em ab initio} forces. We speculate that the strange polarisabilities are an effort 
by our parameterization program to compensate for those other electronic effects which 
are not present in the mathematical form of the potential. It is remarkable, and further 
evidence of the power of the force-fitting approach, that the parameteriser succeeds in 
finding values for the charges and polarisabilities of the atoms that perform so well.

\subsection{Thermal expansion}
Figure 3 compares the dependence of the lattice parameters of the 
rutile structure on temperature with experiment\cite{krishna,meagher}. 
Once again, the agreement is very good, indicating that the anharmonicity 
of the potential energy surface is well reproduced by our force field. 
The results reported in figure 3 were from long ($> 100$ ps) MD simulations of a 
$6\times 6\times 8$ supercell, which contains $1728$ atoms. A time step of $0.723$ 
fs was used and temperature was controlled with a Nose-Hoover thermostat\cite{nose}. 
While atoms moved according to Newton's equations (using a Verlet algorithm), 
steepest descent was performed on the cell degrees of freedom. We verified that 
this approach gave almost identical results to performing Parrinello-Rahman constant 
pressure simulations\cite{pr} and taking averages over the trajectory of the 
lattice constants. However, the latter simulations are more time consuming and more difficult to control.
We caution that our simulations treat the ions purely classically and that quantum effects would 
flatten out these curves at low temperatures.

\begin{figure}
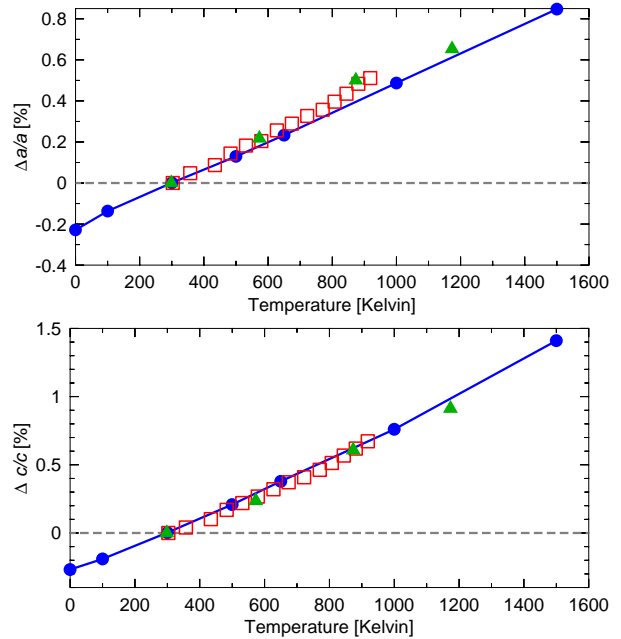

\epsfig{figure=figure3a.ps,width=8.0cm}
\epsfig{figure=figure3b.ps,width=8.0cm}
\caption{Temperature dependence of the lengths of the rutile lattice 
vectors, expressed as percentage differences with respect to their 
room temperature values, i.e. $\Delta a(T) = a(T)-a(300\text{ K})$. 
Boxes and filled triangles refer to the experimental data of Refs.
 \onlinecite{krishna} and \onlinecite{meagher}, respectively. Blue dots are the results of MD simulations with our new force field.}
\label{fig:thermal}
\end{figure}

\subsection{Thermodynamic properties}
\begin{figure}
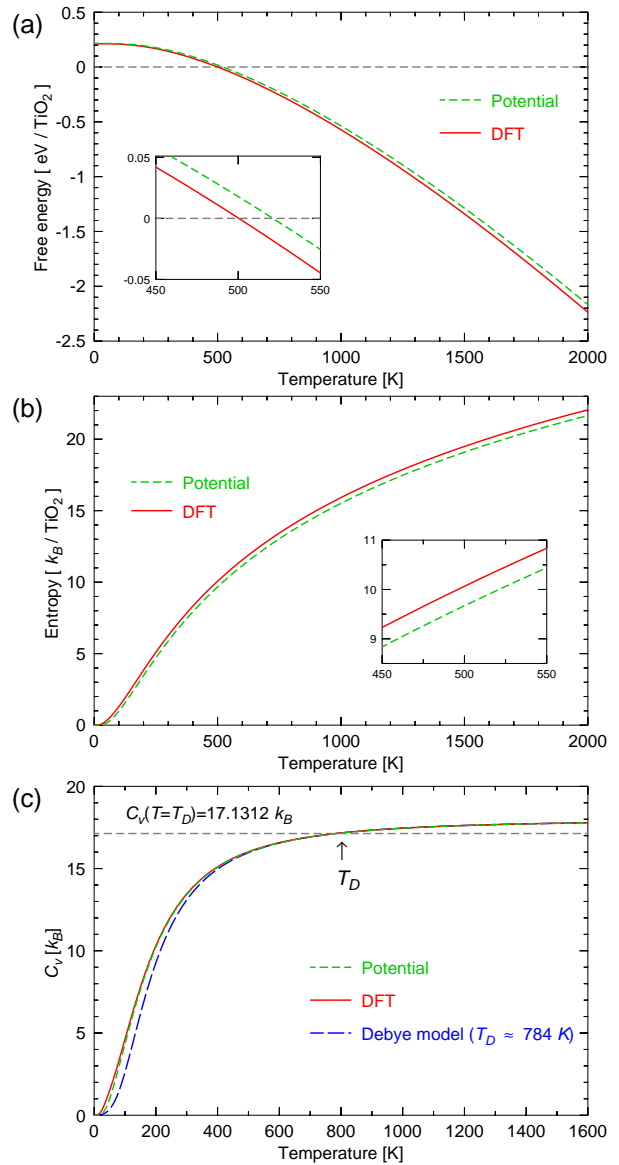

\epsfig{figure=figure4a.ps,width=8.0cm}
\epsfig{figure=figure4b.ps,width=8.0cm}
\epsfig{figure=figure4c.ps,width=8.0cm}
\caption{Thermodynamic properties of rutile TiO$_2$ derived from phonon properties:
       (a) free energy $F$; (b) entropy $S$; (c) specific heat $C_v$ under constant volume}
\label{fig:thermo}
\end{figure}

Thermodynamic properties of rutile TiO$_2$, such as free energy, entropy, and 
phonon specific heat under constant volume were evaluated in the quasi-harmonic 
approximation from the phonons using the FROPHO package \cite{togo} and the results are 
given in Figure \ref{fig:thermo}. We compare the results from DFT calculations of the phonons and 
calculations of the phonons with our force field. In both cases, phonons were calculated in 
$4\times4\times4$ supercell.
As illustrated in Figure \ref{fig:thermo}(a), the temperature dependence of the free energy 
calculated from the new force field is almost the same as that from our {\em ab initio} calculation. 
An enlargement of the free energy curve is also superimposed in Figure \ref{fig:thermo}(a) in order to see the 
difference between these two results more clearly.  The entropy of rutile 
TiO$_2$ is given in Figure \ref{fig:thermo}(b). Again, the new force field reproduces the {\em ab initio}
calculation extremely well. 

Figure \ref{fig:thermo}(c) shows that the specific heat under constant volume ($C_v$)
from the new classical force field is also in excellent agreement with
our {\em ab initio} calculations.
Within the Debye model \cite{debye}, $C_v(T)$ can be expressed as
\begin{equation}
C_v = 9 N k_B \bigg(\frac{T}{T_D}\bigg)^3 \int_0^{T_D/T} \frac{x^4 e^x}{(e^x-1)^2 }dx
\end{equation}
where $N$ is the number of atoms per cell (6 for rutile TiO$_2$), 
$k_B$ is the Boltzmann constant, and $T_D$ is the Debye temperature. 
Using this expression, the specific heat at $T_D$ can be computed as
\begin{equation}
C_v(T=T_D) = 9Nk_B\int_0^1 \frac{x^4e^x}{(e^x-1)^2}dx = 17.1312k_B
\end{equation}
By comparing this value with our calculation, we can obtain the Debye temperature $T_D$.
We find that our {\em ab initio} calculations predict $T_D=781.6$ K 
while our new force field predicts $T_D=786.1$ K. 
These values are very close to the experimental Debye temperature of $778.3$ K \cite{porto}. 

\section{Conclusions}
A new classical force field for TiO$_2$ has been proposed that 
has been parameterized using forces, stresses, and energies 
extracted from {\em ab initio} molecular dynamics simulations. 
The potential can not only predict the structural properties of 
TiO$_2$ in three crystal structures, namely, rutile, brookite, and 
anatase, but also gives the right ground state and the correct 
energy sequence of these three structures and predicts their bulk 
moduli with an accuracy comparable to DFT. 
While the force field has been constructed with rutile in mind, 
and fit to {\em ab initio} data of the rutile structure, it seems 
likely to be reasonably transferable to different structures.
However, the most obvious deficiency of the force-field is that it overestimates energy
differences between these polymorphs. This, and the unphysical polarisabilities of the
ions, suggest that our functional form may
lack ingredients necessary to describe some electronic effects that can play important
roles in the energetics of TiO$_2$. Quadrupole
polarisation of ions and dynamical charge transfer between ions are two effects that
may have an important role to play.

Concerning the vibrational properties, the new potential can 
reproduce most features of {\em ab initio} and experimental results, 
and is a substantial improvement over the widely-used MA model for TiO$_2$. 
Furthermore, the potential provides an excellent description of the 
temperature dependence of the rutile lattice constants. 
The good description of vibrational properties from the proposed force field
is further confirmed by thermodynamic quantities calculated from the phonon properties.
The calculated free energy, entropy, specific heat under constant volume, 
and Debye temperature via the new classical force field are in excellent 
agreement with those from {\em ab initio} calculations. 

Overall, our results indicate that the force field presented 
provides a very accurate description of atomic interactions 
in rutile TiO$_2$. 
The force field can be applied with some confidence to studies of bulk properties of titania. While further testing is 
necessary, we can also hope that it provides an improved description of surface properties. If this is the case, it can be 
used to study, not only surfaces, but also nanocrystals and nanowires, all of which have 
huge technological importance for their catalytic and optical properties.

We stress that no experimental data, other than the 
rutile crystal symmetry, has been used to parameterize this 
potential, therefore, we can be confident that its predictive 
capability is based on a robust description of the ions' potential-energy surface. 

\section{Acknowledgments}
X. J. H. acknowledges useful discussions with H. Schober.
P.T. acknowledges useful discussions with M. W. Finnis and N. M. Harrison.
L.B. acknowledges support from the EU within the Marie Curie Actions for Human Resources and Mobility.
P.T. acknowledges support from the European Commission within the Marie Curie Support for Training and Development
of Researchers Programme under Contract No. MIRG-CT-2007-208858.


\begin{references}

\bibitem{jeong1}  
D. S. Jeong, H. Schroeder, and R. Waser, Phys. Rev. B {\bf 79}, 195317 (2009).

\bibitem{jeong2}  
D. S. Jeong, H. Schroeder, and R. Waser, Electrochem. Solid-State Lett. {\bf 10}, G51 (2007).

\bibitem{shima}  
H. Shima, N. Zhong, and H. Akinaga, Appl. Phys. Lett. {\bf 94}, 082905 (2009).

\bibitem{regan}  
B. O'Regan and M. Gr\"{a}tzel, Nature {\bf 353}, 737 (1991).

\bibitem{khan}  
S. U. M. Khan, M. Al-Shahry, W. B. Ingler Jr., Science {\bf 297}, 2243 (2002).

\bibitem{du}  
Y. Du, N. A. Deskins, Z. Zhang, Z. Dohnalek, M. Dupuis, 
and I. Lyubinetsky, Phys. Rev. Lett. {\bf 102}, 096102 (2009).

\bibitem{matsui}  
M. Matsui, and M. Akaogi, Mol. Simu. {\bf 6}, 239 (1991).

\bibitem{swamy}  
V. Swamy, J. D. Gale, and L. S. Dubrovinsky, J. Phys. Chem. Solids {\bf 62}, 887 (2001).

\bibitem{ogata}  
S. Ogata, H. lyetomi, K. Tsuruta, F. Shimojo, R. K. Kalia, 
A. Nakano, and P. Vashishta, J. Appl. Phys. {\bf 86}, 3036 (1999).

\bibitem{collins}   
D. R. Collins, W. Smith, Council for the Central Laboratory of the 
Research Councils Tech. Rep. DL-TR-96-001, 1996.

\bibitem{hallil}  
A. Hallil, R. T\'etot, F. Berthier, I. Braems, and J. Creuze, Phys. Rev. B {\bf 73}, 165406 (2006).

\bibitem{kerisit}
S. Kerisit, N. A. Deskins, K. M. Rosso, and M. Dupuis, J. Phys. Chem. C {\bf 112}, 7678 (2008).

\bibitem{thomas}
B. S. Thomas, N. A. Marks, and B. D. Begg, Phys. Rev. B {\bf 69}, 144122 (2004).

\bibitem{tetot}
R. T\'etot, A. Hallil, J. Creuze, and I. Braems, Europhys. Lett. {\bf 83} 40001, (2008).


\bibitem{ercolessi}
F. Ercolessi and J. B. Adams, Europhys. Lett. {\bf 26}, 583 (1994).

\bibitem{ts}  
P. Tangney and S. Scandolo, J. Chem. Phys. {\bf 117}, 8898 (2002).

\bibitem{ts-mgo}   
P. Tangney and S. Scandolo, J. Chem. Phys. {\bf 119}, 9673 (2003).

\bibitem{brommer}  
P. Brommer and F. G\"{a}hler, Modelling Simul. Mater. Sci. Eng. {\bf 15}, 295 (2007).

\bibitem{traylor}  
J. G. Traylor, H. G. Smith, R. M. Nicklow, and M. K. Wilkinson, Phys. Rev. B {\bf 3}, 3457 (1971).

\bibitem{porto}  
S. P. S. Porto, P. A. Fleury, T. C. Damen, Phys. Rev. {\bf 154}, 522 (1962).

\bibitem{spitzer}  
W. G. Spitzer, R. C. Miller, D. A. Kleinman, and L. E. Howarth, Phys. Rev. {\bf 126}, 1710 (1962).

\bibitem{montanari}  
B. Montanari and N. M. Harrison, Chem. Phys. Lett. {\bf 364}, 528 (2002).

\bibitem{lee}  
C. Lee, P. Ghosez, and X. Gonze, Phys. Rev. B {\bf 50}, 13379 (1994).

\bibitem{sikora}   
R. Sikora, J. Phys. Chem. Sol. {\bf 66}, 1069 (2005).

\bibitem{aguado}  
A. Aguado and P. A. Madden, Phys. Rev. B {\bf 70}, 245103 (2004).

\bibitem{wilson-madden1} 
M. Wilson and P. A. Madden, J. Phys: Cond. Matt. {\bf 5}, 2687 (1993).

\bibitem{rowley}  
A. J. Rowley, P. Jemmer, M. Wilson, and P. A. Madden, J. Chem. Phys. {\bf 108}, 10209 (1998).

\bibitem{vasp1}  
G. Kresse and J. Hafner, Phys. Rev. B {\bf 47}, 558 (1993).

\bibitem{vasp2}  
G. Kresse and J. Furthm\"uller, Phys. Rev. B {\bf 54}, 11169 (1996).

\bibitem{vasp3}  
G. Kresse and D. Joubert, Phys. Rev. B {\bf 59}, 1758 (1999).

\bibitem{verlet}
L. Verlet, Phys. Rev. {\bf 159}, 98 (1967); Phys. Rev. {\bf 165}, 201 (1967).

\bibitem{nose}
S. Nos\'e, J. Chem. Phys. {\bf 81}, 511 (1984); S. Nose, Mol. Phys. {\bf 52}, 255 (1984); 
W. G. Hoover, Phys. Rev. A {\bf 31}, 1695 (1985).

\bibitem{simann}  
S. Kirkpatrick, C. D. Gelatt, and M. P. Vecchi, Science {\bf 220}, 671 (1983).

\bibitem{numrec}  
W. H. Press, W. T. Vetterling, B. P. Flannery, and S. A. Teukolsky, 
Numerical Recipes in Fortran 77 and Fortran 90: The Art of Scientific and Parallel Computing, 2nd ed.  
(Cambridge University Press, Cambridge, 1996).

\bibitem{abrahams}  
S. C. Abrahams, J. L.Bernstein, J. Chem. Phys. {\bf 55}, 3206 (1971)

\bibitem{burdett}  
J. K. Burdett, T. Hughbanks, G. J. Miller, J. W. Richardson, and J. V. Smith, 
J. Am. Chem. Soc. {\bf 109}, 3639 (1987).

\bibitem{meagher}  
E. P. Meagher and G. A. Lager, Canad. Mineral. {\bf 17}, 77 (1979).

\bibitem{mitsuhashi}  
T. Mitsuhashi and O. J. Kleppa, J. Amer. Ceram. Soc. {\bf 62}, 356 (1979).

\bibitem{navrotsky}
A. Navrotsky and O. J. Kleppa, J. Amer. Ceram. Soc. {\bf 50}, 626 (1967); A. Navrotsky, J. C. Jamieson, 
and O. J. Kleppa, Science {\bf 158} 388 (1967).

\bibitem{labat}  
F. Labat, P. Baranek, C. Domain, C. Minot, and C. Adamo, J. Chem. Phys. {\bf 126}, 154703 (2007). 

\bibitem{pbe}
J. P. Perdew, K. Burke, and M. Ernzerhof, Phys. Rev. Lett. {\bf 77}, 3865 (1996).

\bibitem{birch}   
F. Birch, Phys. Rev. {\bf 71}, 809 (1947).

\bibitem{ming}  
L. Ming and M. H. Manghnani, J. Geophys. Res. {\bf 84}, 4777 (1979).

\bibitem{gerward}  
L. Gerward and J. S. Olsen, J. Appl. Crystallogr. {\bf 30}, 259 (1997).

\bibitem{togo}  
A. Togo, FROPHO: Phonon analyser for periodic boundary condition materials; 
A. Togo, F. Oba, and I. Tanaka, Phys. Rev. B {\bf 78}, 134106 (2008). 

\bibitem{born-huang}
M. Born and K. Huang, {\em Dynamical Theory of Crystal Lattices}, Oxford Univ. Press, 1954.

\bibitem{debye}  
P. Debye, Ann. Phys. {\bf 39}, 789 (1912).

\bibitem{krishna}   
K. V. Krishna Rao, S. V. Nagender Naidu, and L. Iyengar, J. Amer. Cer. Soc. {\bf 53}, 124 (1970).

\bibitem{liang}  
Y. Liang, C. R. Miranda, and S. Scandolo, J. Chem. Phys. {\bf 125}, 194524 (2006).

\bibitem{rignanese}  
G.-M. Rignanese, X. Rocquefelte, X. Gonze, 
and A. Pasquarello, Int. J. Quant. Chem. {\bf 101}, 793 (2005).

\bibitem{pr}
M. Parrinello and A. Rahman, Phys. Rev. Lett. {\bf 45}, 1196 (1980).
\end{references}
\end{document}